# Spin Hall Conductivity in the impure two-dimensional Rashba s-wave superconductor

M Biderang, H Yavari[*]

Department of Physics, University of Isfahan, HezarJarib St, Isfahan 81746, Iran

**Abstract.** Based on the Kubo formula approach, the spin Hall conductivity (SHC) of a two-dimensional (2D) Rashba s-wave superconductor in the presence of nonmagnetic impurities is calculated. We will show that by increasing the superconducting gap, the SHC decreases monotonically to zero, while by decreasing the concentration of impurities at zero gap, the SHC closes to the clean limit universal value $-\frac{e}{8\pi}$. As a function of the impurity relaxation rate $\tau$ at $T_c = 0.1$ and $\gamma = 0.01$ ($\gamma$ is the spin-orbit coupling in unit of $eV.m$), we will show that in the dirty limit ($\tau \to 0$) the SHC vanishes, and by increasing the relaxation time ($\tau \to \infty$) the SHC depends on the value of superconducting gap ($\Delta = 1.76 T_c \sqrt{1 - \frac{T}{T_c}}$), is changed from zero for full gap to $-\frac{e}{8\pi}$ in zero gap. At low temperatures, the SHC goes to zero exponentially and near the critical temperature depending on the concentration of the scattering centers, the SHC will tend to the value of normal state. We will also show that the SHC is independent of spin-orbit coupling ($\gamma$) in the clean limit.

## 1. Introduction

Spin Hall effect (SHE) is an intriguing phenomenon, in which a longitudinal electric field produces a transverse spin current due to spin-orbit interactions. This topic has attracted much recent theoretical and experimental attention [1- 6]. Spin-orbit coupling (SOC) as an important interactions in quantum condensed matter physics, is often revealed by a linear coupling between the internal (spin) state of a particle and its motional (orbital) degree of freedom, which fastens the spin state of a particle to its direction of motion. The correlation due to this interaction becomes essential in condensed matter phenomena, such as topological insulator [7, 8], spin Hall effect [9, 10], and spintronics [11].

---

[*]Corresponding author, Fax: (+98)3137922409
E-mail address: h.yavari@sci.ui.ac.ir

SHE is one of the most important aspects of the existence of spin orbit interaction (SOI) in the electron transport systems, which can occur in metals, semiconductors and superconductors. Theoretically, the Spin Hall effect can be described in terms of two completely distinct physical mechanisms: intrinsic spin Hall effect (ISHE) and extrinsic spin Hall effect (ESHE). The latter is controlled by the SOI with impurities, and the former could be due to structure inversion asymmetry of the confining potential (leading to Rashba coupling) or due to bulk inversion asymmetry (allowing for Dresselhaus coupling) [12]. In the recent years, many research attempts have focused on the SHE in metals, which show a large (SHC), e.g. SHC in Pt is $200\hbar e^{-1}\Omega^{-1}cm^{-1}$ and those in Nb and Mo are large negative numbers [13].

Application of Superconductivity to spin transports often causes novel phenomena to spintronics. Theoretical predictions show that superconductivity may increase the SHE due to the enhancement in the resistance of superconducting quasiparticles which mediate spin transport in superconductors [14, 15]. The main technique to control the spin state in these devices is an electric filed which has the capability to control the flow of electrons in nanometer scale. Since a dc electric filed cannot exist in the bulk of a superconductor, it is reasonable to ask how to produce an electric current in a superconducting system to emerge SHE. SHE in superconductors is mediated by quasiparticles instead of electrons [16]. These quasiparticles are superposition of electron-like and hole-like excitations [17] and can be injected to superconductors via superconducting tunneling junctions, like a ferromagnet-superconductor-ferromagnet which leads to charge imbalance effect that is compensated by Cooper pair charge to maintain the neutrality of the system [18].

The SHE has been studied using different techniques, including Boltzmann equation, Kubo formula and Keldysh formalism. The Keldysh formalism, however, when applied to SHE, leads to cumbersome relationships. Recently, many attempts have been devoted to investigate the $\sigma_{SH}$ in a normal system. Universal SHC in a clean two-dimensional electron gas (2DEG) was proposed [1, 4], and later using a Boltzmann approach [19, 20] as well as diagrammatic methods like the Kubo formalism in the presence of non-magnetic impurities, it was shown that this universality is absent [21- 23]. The SHC of a disordered two-dimensional Rashba electron gas within the self-consistent Born approximation for arbitrary values of the electron density have been calculated [24]. Using a simple Hamiltonian with Rashba spin-orbit coupling and exchange interactions, the intrinsic SHC in a quantum well semiconductor doped with magnetic impurities, has been computed analytically and it was demonstrated that using the appropriate order of limits, one recovers the intrinsic universal value [25]. The SHC in a 2D Rashba electron gas with electron-electron and electron-phonon interactions in the presence of magnetic and non-magnetic impurities has also been studied [26]. For superconductors the ISHE in the s-wave superconductor with Rashba spin-orbit interaction at finite temperatures has studied [27]. The possibility of SHE in a variety of the iron-based superconducting materials have been explored theoretically and it was found that the origin of the large SHE comes from the huge contribution of electronic states in the vicinity of the SOC- induced gap at the Dirac points that lie close to the Fermi level in the heavily hole doped regime [28].

In this Letter we confine ourselves to the quasiparticle contribution to SHC in 2D Rashba s-wave superconductor in the presence of nonmagnetic impurities. First, we derive a general expression for SHC using the Kubo formula approach and then numerically, the influence of superconducting gap, strength of spin-orbit coupling, concentration of non-magnetic impurity and temperature on the SHC are presented (throughout the paper we use the units $k_B = \hbar = 1$).

## 2. Description of the model

The two dimensional Rashba electron gas can be described by the following Hamiltonian in the band space representation [29],

$$H = \sum_{\vec{k}\sigma} \xi_{\vec{k}} c^\dagger_{\vec{k}\sigma} c_{\vec{k}\sigma} + \sum_{\vec{k}\sigma\sigma'} \vec{\gamma}(\vec{k})\cdot\vec{\sigma}_{\sigma\sigma'} c^\dagger_{\vec{k}\sigma} c_{\vec{k}\sigma'} + \tfrac{1}{2}\sum_{\vec{k}\sigma\sigma'}(\Delta_{\sigma\sigma'}(\vec{k}) c^\dagger_{\vec{k}\sigma} c^\dagger_{-\vec{k}\sigma'} + h.c.) + U_{imp}\sum_{\substack{\vec{R}_j,\sigma \\ \vec{k},\vec{q}}} e^{i\vec{q}\cdot\vec{R}_j} c^\dagger_{\vec{k}\sigma} c_{\vec{k}-\vec{q},\sigma} \quad (1)$$

where $\sigma_i (i = x, y, z)$ is the Pauli matrix, $c^\dagger_{\vec{k}\sigma}(c_{\vec{k}\sigma})$ is the creation (annihilation) operator for an electron with momentum $\vec{k}$ and spin $\sigma$, $\xi_{\vec{k}} = \dfrac{\vec{k}^2}{2m} - \mu$ is the energy dispersion, $\mu$ is chemical potential, $\Delta$ is the superconducting gap which hereafter we assume that $0 \leq \Delta \ll \mu$, and $\vec{\gamma}(\vec{k}) = \gamma|\vec{k}|(-\sin\varphi, \cos\varphi)$ is an odd function which denotes the Rashba spin-orbit interaction ($\gamma$ is the SOI coupling and $\varphi$ is the polar angle). We also assumed that the spin-orbit splitting, $\Delta_{SO} = 2\gamma k_F$, is much smaller than $\mu$. With this assumption, the triplet pairing induced by Rashba SOI becomes negligible [30- 32]. $U_{imp}$ is the non-magnetic impurity potential which we consider rather short-ranged such that s-wave scattering is dominant.

The Kubo formula for the SHC is given by [33],

$$\sigma_{SHE} = -\lim_{\omega \to 0}\frac{\text{Im}[K^{ret}_{SHE}(\omega)]}{\omega} \quad (2)$$

here, $K^{ret}_{SHE}$, the retarded correlation function for spin and charge current, in the Matsubara representation is,

$$K_{SHE}(i\omega_n) = <Tr[J_y^{S_z}(\vec{k})\Im(\vec{k},i\omega_n+ip_m)\Gamma_c^{n,x}(\vec{k};i\omega_n+ip_m,ip_m)\Im(\vec{k},ip_m) +$$
$$+J_y^{S_z}(\vec{k})F(\vec{k},i\omega_n+ip_m)\Gamma_c^{a,x}(\vec{k};i\omega_n+ip_m,ip_m)F(\vec{k},ip_m)]> \quad (3)$$

where $Tr[...] = \frac{1}{L^2}\sum_{ip_m}\int\frac{d^2k}{(2\pi)^2}Tr_s(...)$, $\omega_n = 2n\pi T$ and $p_m = (2m+1)\pi T$ are bosonic and fermionic Matsubara frequencies respectively, $J_y^{S_z} = \frac{1}{2e}\{J_y^C, S_z\} = \frac{k_y\sigma_z}{2m}$ is the spin current operator in y direction for spin polarized along z, and $J_x^C(\vec{k}) = -\frac{\delta H(\vec{k}+e\vec{A})}{\delta A_y} = e(\frac{k_x}{m}\hat{I} + \gamma\sigma_y)$ is the bare charge current operator for particle channel ($e\gamma\sigma_y$ is the anomalous velocity essential for SHE). The normal and anomalous Matsubara Green's functions ($\Im(\vec{k},ik_m)$, $F(\vec{k},ik_m)$) for a superconductor in spinor representation respectively can be written as,

$$\Im(\vec{k},ik_m) = \sum_{s=\pm 1}\hat{\Pi}_s(\vec{k})\Im_s(k,ik_m) \quad (4)$$

$$F(\vec{k},ik_m) = \sum_{s=\pm 1}\hat{\Pi}_s(\vec{k})F_s(k,ik_m) \quad (5)$$

here, $\hat{\Pi}_s(\vec{k}) = \frac{1}{2}[\hat{I} + s(\hat{k}_x\sigma_y - \hat{k}_y\sigma_x)]$ is the band projection operator. The normal and anomalous Matsubara Green's functions in the helicity basis respectively are,

$$\Im_s(k,ik_m) = \frac{ik_m - \Sigma_{imp}(ik_m) + \xi_{k,s}}{(ik_m - \Sigma_{imp}(ik_m) + \xi_{k,s})(ik_m - \Sigma_{imp}(ik_m) - \xi_{k,s}) + |\Delta_{k,s}|^2} \quad (6)$$

$$F_s(k,ik_m) = \frac{-\Delta_{k,s}}{(ik_m - \Sigma_{imp}(ik_m) + \xi_{k,s})(ik_m - \Sigma_{imp}(ik_m) - \xi_{k,s}) + |\Delta_{k,s}|^2} \quad (7)$$

where, $\xi_{k,s} = \xi_k + s\gamma k$ is the energy dispersion in normal state and $\Sigma_{imp}$ is the self-energy due to the impurities where its imaginary part is related to the relaxation rate $\tau_{imp}^{-1} = 2\,\text{Im}|\Sigma_{imp}| = 2\pi n_{imp}N_0 U_{imp}^2$ ($n_{imp}$ is the concentration of impurities and $N_0 = \frac{m}{2\pi}$ is the density of states per spin direction in normal state). The normal (n) and anomalous (a) parts of the vertex function are defined respectively as,

$$\Gamma_c^{x,n}(\vec{k};iv_m+i\omega_n,iv_m) =$$
$$J_x^C(\vec{k})+\frac{1}{2\pi\tau N_0}\left\langle Tr[\Im(\vec{k}',iv_m+i\omega_n)\Gamma_c^{x,n}(\vec{k}';iv_m+i\omega_n,iv_m)\Im(\vec{k}',iv_m)]\right\rangle \quad (8)$$

$$\Gamma_c^{x,a}(\vec{k};iv_m+i\omega_n,iv_m) =$$
$$J_x^C(\vec{k})+\frac{1}{2\pi\tau N_0}\left\langle Tr[F(\vec{k}',iv_m+i\omega_n)\Gamma_c^{x,a}(\vec{k}';iv_m+i\omega_n,iv_m)F(\vec{k}',iv_m)]\right\rangle \quad (9)$$

After some algebra, Eqs. (8) and (9) can be written as,

$$\Gamma_c^{x,n}(k;iv_m+i\omega_n,iv_m) = \frac{e}{m}k_x + e\gamma\sigma_y\Gamma_n(iv_m+i\omega_n,iv_m) \quad (10)$$

$$\Gamma_c^{x,a}(k;iv_m+i\omega_n,iv_m) = \frac{e}{m}k_x + e\gamma\sigma_y\Gamma_a(iv_m+i\omega_n,iv_m) \quad (11)$$

Substituting Eqs.(10) and (11) into Eq.(3) we get,

$$K_{SHE}(i\omega_n) = 2i\frac{e\gamma}{m}T\sum_m\{\Gamma_n(i\omega_n+iv_m,iv_m)B_{1,n}(i\omega_n+iv_m,iv_m)+\Gamma_a(i\omega_n+iv_m,iv_m)B_{1,a}(i\omega_n+iv_m,iv_m)\} \quad (12)$$

here

$$\Gamma_n(iv_m+i\omega_n,iv_m) = 1+\sum_{m'}\frac{1}{2\pi\tau N_0 k_0}[B_{2,n}(iv_{m'}+i\omega_n,iv_{m'})+k_0 B_{3,n}(iv_{m'}+i\omega_n,iv_{m'})\Gamma_n(iv_{m'}+i\omega_n,iv_{m'})] \quad (13)$$

$$\Gamma_a(iv_m+i\omega_n,iv_m) = 1+\sum_{m'}\frac{1}{2\pi\tau N_0 k_0}[B_{2,a}(iv_{m'}+i\omega_n,iv_{m'})+k_0 B_{3,a}(iv_{m'}+i\omega_n,iv_{m'})\Gamma_a(iv_{m'}+i\omega_n,iv_{m'})] \quad (14)$$

The bubble functions $B_{i,n}$ and $B_{i,a}$ ($i=1,2,3$) in Eqs.(12)-(14) are defined as,

$$B_{1,n}(i\omega_n+iv_m,iv_m) = \sum_s s\int\frac{d^2k}{(2\pi)^2}k\sin^2\varphi[\Im_{-s}(k,i\omega_n+iv_m)\Im_s(k,iv_m)] \quad (15)$$

$$B_{2,n}(i\omega_n+iv_m,iv_m) = \sum_s 4s\int\frac{d^2k}{(2\pi)^2}k\cos^2\varphi[\Im_s(k,i\omega_n+iv_m)\Im_s(k,iv_m)] \quad (16)$$

$$B_{3,n}(i\omega_n + i\nu_m, i\nu_m) = \sum_{s,s'} 2\int \frac{d^2k}{(2\pi)^2}(1+ss'\cos(2\varphi))[\Im_s(k,i\omega_n+i\nu_m)\Im_{s'}(k,i\nu_m)] \quad (17)$$

$$B_{1,a}(i\omega_n + i\nu_m, i\nu_m) = \sum_s s\int \frac{d^2k}{(2\pi)^2} k\sin^2\varphi[F_{-s}(k,i\omega_n+i\nu_m)F_s(k,i\nu_m)] \quad (18)$$

$$B_{2,a}(i\omega_n + i\nu_m, i\nu_m) = \sum_s 4s\int \frac{d^2k}{(2\pi)^2} k\cos^2\varphi[F_s(k,i\omega_n+i\nu_m)F_s(k,i\nu_m)] \quad (19)$$

$$B_{3,a}(i\omega_n + i\nu_m, i\nu_m) = \sum_{s,s'} 2\int \frac{d^2k}{(2\pi)^2}(1+ss'\cos(2\varphi))[F_s(k,i\omega_n+i\nu_m)F_{s'}(k,i\nu_m)] \quad (20)$$

Eq. (12) can be simplified as,

$$K_{SHE}(i\omega_n) = T\sum_m [\kappa_n(i\omega_n+i\nu_m, i\nu_m) + \kappa_a(i\omega_n+i\nu_m, i\nu_m)] \quad (21)$$

where

$$\kappa_n(i\omega_n+i\nu_m, i\nu_m) = 2i\frac{e\gamma}{m}\Gamma_n(i\omega_n+i\nu_m, i\nu_m)B_{1,n}(i\omega_n+i\nu_m, i\nu_m) \quad (22)$$

$$\kappa_a(i\omega_n+i\nu_m, i\nu_m) = 2i\frac{e\gamma}{m}\Gamma_a(i\omega_n+i\nu_m, i\nu_m)B_{1,a}(i\omega_n+i\nu_m, i\nu_m) \quad (23)$$

After doing the Matsubara frequency summation in Eq. (21) and performing the analytical continuation, $i\omega_n \to \omega + i\delta$, we obtain,

$$K_{SHE}^{ret}(\omega) = -\int_{-\infty}^{\infty} \frac{d\varepsilon}{2\pi i}(f(\varepsilon+\omega)-f(\varepsilon))[\kappa_n(\varepsilon+\omega+i\delta,\varepsilon-i\delta)+\kappa_a(\varepsilon+\omega+i\delta,\varepsilon-i\delta)] -$$
$$-\int_{-\infty}^{\infty} \frac{d\varepsilon}{2\pi i} f(\varepsilon)[\kappa_n(\varepsilon+\omega+i\delta,\varepsilon+i\delta)+\kappa_a(\varepsilon+\omega+i\delta,\varepsilon+i\delta)] + \quad (24)$$
$$+\int_{-\infty}^{\infty} \frac{d\varepsilon}{2\pi i} f(\varepsilon+\omega)[\kappa_n(\varepsilon+\omega-i\delta,\varepsilon-i\delta)+\kappa_a(\varepsilon+\omega-i\delta,\varepsilon-i\delta)]$$

When we calculate the spin-Hall conductivity via Eqs.(24), and (2) it is evident that the resulting $\sigma_{SHE}$, according to different combinations of retarded (R) and advanced (A) normal and anomalous Green's functions, will be written as the sum of contributions $\sigma_{SHE}^{RA}$, $\sigma_{SHE}^{RR}$, and $\sigma_{SHE}^{AA}$ respectively, defined as the first, second, and third terms in the right hand side of Eq.(24).The second and third terms which come from processes away from the Fermi surface can be neglected in compared with the first term in the large Fermi energy limit and finally the SHC can be written as,

$$\sigma_{SHE} = \frac{1}{2\pi i}\int_{-\infty}^{\infty} d\varepsilon (\frac{\partial f}{\partial \omega})_{\omega=\sqrt{\varepsilon^2+\Delta^2}}[\kappa_n(\varepsilon+\omega+i\delta,\varepsilon-i\delta)+\kappa_a(\varepsilon+\omega+i\delta,\varepsilon-i\delta)] \qquad (25)$$

Eq. (25) is a general expression which contains all effect (Temperature, superconducting gap, impurity relaxation rate, and spin-orbit coupling) on the SHC and in the following section we will evaluate it numerically.

## 3. The variation of SHC with superconducting gap, concentration of non-magnetic impurities, temperature and spin-orbit coupling.

In this section, by using Eq. (25) the effects of superconductivity gap, concentration of non-magnetic impurities, temperature and spin-orbit coupling constant on the SHC are discussed numerically.

We, first, consider the effect of superconducting gap on the SHC. Figure 1 shows the variation of SHC as a function of superconducting gap for several values of the impurity relaxation time ranging from $\tau_{imp}=10^1$ to $\tau_{imp}=10^4$. As the superconducting gap increases, the SHC decreases monotonically to zero for $\Delta=0.05$, in consist with the quasiparticle contribution at the Fermi level reported in Ref. [27], while by increasing $\tau$ at zero gap, the SHC closes to its clean limit universal value $-\frac{e}{8\pi}$.

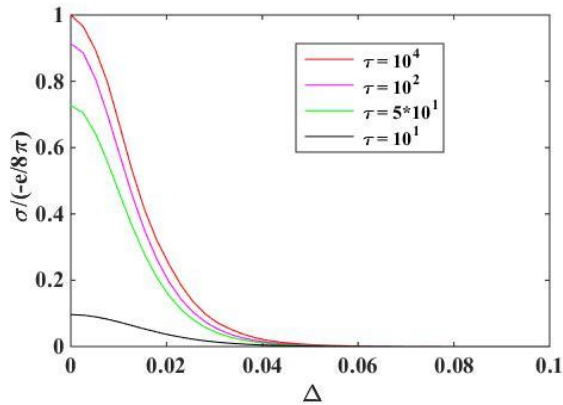

Figure 1: The SHC as a function of $\Delta$ for different values of impurity relaxation time

In figure 2, we present SHC as a function of impurity relaxation time for different values of temperature and superconductivity gap at $T_c = 0.1$ and $\gamma = 0.01$. We observe that in the dirty limit ($\tau \to 0$) the SHC vanishes, and by increasing the relaxation time ($\tau \to \infty$) depends on the value of the temperature and superconducting gap ($\Delta = 1.76 T_c \sqrt{1 - \frac{T}{T_c}}$), the SHC is changed from zero for full gap to $-\frac{e}{8\pi}$ in the absence of the gap. At $T = T_c = 0.1$ ($\Delta = 0$), the SHC is reduced to $-\frac{e}{8\pi}$ as the impurity relaxation time increases (concentration of scattering centers decreases) in agreement with the result shown in Fig.1 of Ref. [34], but decreases toward zeros as temperature departs from the critical value (below the critical temperature) due to enhancement of the superconducting gap.

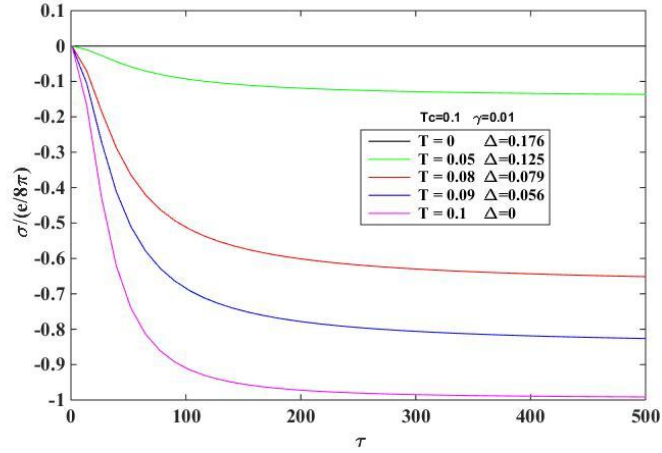

Figure 2: The SHC as a function of impurity relaxation time for different temperatures and superconducting gaps

In figure 3 we report the effect of temperature ($t = \frac{T}{T_c}$) on the SHC of s-wave superconductor for different value of the relaxation time (Fig. 3 (a) - (d)). At low temperatures ($t \to 0$), the SHC goes to zero exponentially, which is in good agreement with the quasiparticle contribution to the SHC in Ref. [28]. As temperature approach to the critical one, ($t \to 1, \Delta \to 0$), depending on the concentration of the scattering centers, the SHC will tend to the value of normal state. For the clean limit, $\tau \to \infty$ this normal value is $-\frac{e}{8\pi}$. According to Refs.[14, 15], the quasiparticles are responsible to carry the current in the superconductors and at zero temperature, there are not such excitations and eventually, there is not any transverse spin current, which leads to zero SHC.

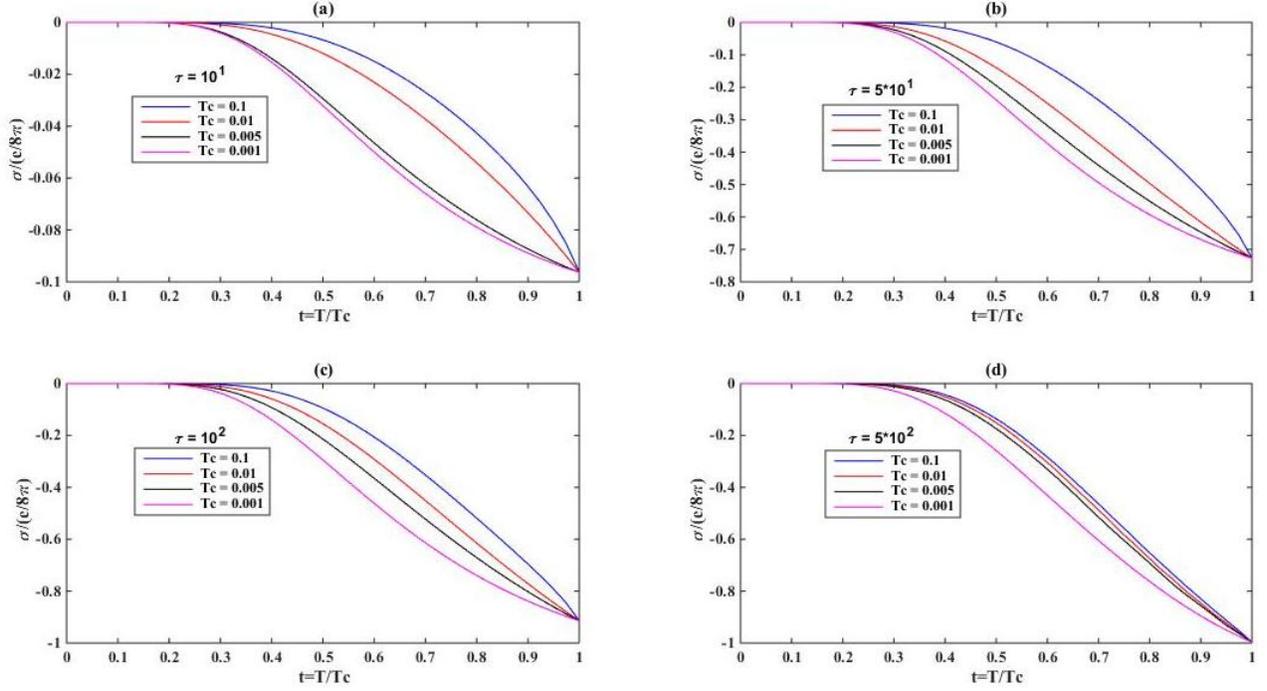

Figure 3: The SHC as a function of $t = \dfrac{T}{T_c}$ for (a): $\tau = 10^1$, (b): $\tau = 5 \times 10^1$, (c): $\tau = 10^2$ and (d): $\tau = 5 \times 10^2$

In figure 4 the effect of spin-orbit coupling on the SHC is represented. In the clean limit, $\tau \to \infty$, the SHC is independent of spin-orbit coupling which is in a good agreement with Ref. [27], but as the relaxation time decreases, in the dirty limit ($\tau \to 0$), the SHC will be affected by $\gamma$.

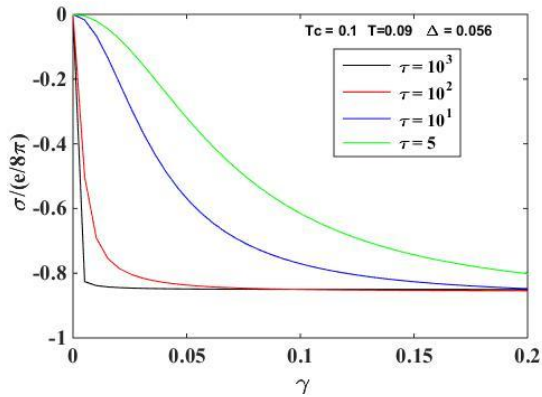

Figure 4: The SHC as a function of γ for different values of impurity relaxation time

**4. Conclusion**

We have calculated the SHC of an impure conventional superconductor with spin-orbit coupling, in the framework of linear response theory. In the weak impurity scattering regime, the expression for the static SHC as a function of the system's parameters (spin-orbit coupling, scattering time, Fermi energy) was obtained and the effect of superconducting gap, impurity concentration, temperature and spin-orbit coupling strength on the SHC have been analyzed numerically. The behavior of the SHC as a function of superconducting gap for different impurity concentrations shows that as the superconducting gap increases, the SHC decreases monotonically to zero for $\Delta = 0.05$, while by increasing the impurity relaxation time $\tau$ at zero gap, the SHC closes to its clean limit universal value $-\frac{e}{8\pi}$.

We also explored the influence of the impurity relaxation time to the SHE and we found that the impurity scattering plays an important role in SHE, leading to a nonzero SHC for arbitrary concentrations. In the dirty limit, the SHC goes to zero and it leads to $-\frac{e}{8\pi}$ in the clean limit.

The temperature-dependent SHC reported here shows that by increasing the temperature, the SHC departs from its zero value at low temperatures and at the critical temperature where the superconducting gap closes to zero will tend to the value of the normal state. For the clean limit, $\tau \to \infty$, this normal value is $-\frac{e}{8\pi}$ and by decreasing the impurity relaxation time, the SHC departs from this universal value.

Finally, the effect of spin-orbit coupling on the SHC was represented for several values of the relaxation time. In the clean limit, $\tau \to \infty$, the SHC is clearly independent of spin-orbit coupling $\gamma$, but as the relaxation time decreases, in the dirty limit, the SHC will be affected by $\gamma$, obviously.

**References**

[1] J. Sinova, D. Culcer, Q Niu, N. A. Sinitsyn, T. Jungwirth, and A. H. MacDonald, Universal Intrinsic Spin Hall Effect *Phys. Rev. Lett.* **92** (2004) 126603.

[2] J. Sinova, S. O. Valenzuela, J. Wunderlich, C. H. Back, and T. Jungwirth, Spin Hall effects, *Rev. Mod. Phys.* **87**(2015) 1213-60.

[3] R. Raimondi, P. Schwab, C. Gorini, and G. Vignale, Spin-orbit interaction in a two-dimensional electron gas: A SU(2) formulation, *Ann. Phys. (Berlin)* **524** (2012)153.

[4] S. Murakami, N. Nagaosa, and S.C. Zhang, Dissipationless Quantum Spin Current at Room Temperature, *Science* **301** (2003)1348-51.

[5] J. Inoue, G. E. W. Bauer, and L. W. Molenkamp, Suppression of the persistent spin Hall current by defect scattering, *Phys. Rev. B* **70** (2004) 041303.

[6] K. Nomura, J. Sinova, N. A. Sinitsyn, and A. H. MacDonald, Dependence of the intrinsic spin-Hall effect on spin-orbit interaction character, *Phys. Rev. B* **72** (2005)165316.

[7] M. Z. Hasan and C. L. Kane, Colloquium: Topological insulators, *Rev. Mod. Phys.* **82**(2010) 3045-67.

[8] X. L. Qi and S.C. Zhang, Topological insulators and superconductors, *Rev. Mod. Phys.* **83** (2011)1057-110.

[9] Y. K. Kato, R. C. Myers, A. C. Gossard, and D. D. Awschalom, Observation of the Spin Hall Effect in Semiconductors *Science* **306** (2004)1910-3.

[10] M. Konig, S. Wiedmann, C. Brune, A. Roth, H. Buhmann, L.W. Molenkamp, X.L Qi, and S.C. Zhang, Quantum Spin Hall Insulator State in HgTe Quantum Wells, *Science* **318**( 2007) 766-70.

[11] I. Zuti, J. Fabian, and S. Das Sarma, Spintronics: Fundamentals and applications, *Rev. Mod. Phys.* **76** (2004)323-410.

[12] G. Vignale, Ten Years of Spin Hall Effect, *J. Supercond. Nov. Magn.* **23** (2010) 3-10.

[13] T. Kimura, Y. Otani, T. Sato, S. Takahashi, and S. Maekawa, Room-Temperature Reversible Spin Hall Effect *Phys. Rev. Lett.* **98** (2007)156601.

[14] S. Takahashi and S. Maekawa, Hall Effect Induced by a Spin-Polarized Current in Superconductors, *Phys. Rev. Lett.* **88** (2002) 116601.

[15] S. Takahashi and S. Maekawa, Spin Current in Metals and Superconductors *J. Phys. Soc. Jpn.* **77** (2008) 031009.

[16] T. Wakamura, H. Akaike, Y. Omori, Y. Niimi, S. Takahashi, A. Fujimaki, S. Maekawa, and Y. Otani, Quasiparticle-mediated spin Hall effect in a superconductor, *Nat. Mater* **14**(2015) 675-8.

[17] M. Tinkham, *Introduction to superconductivity*, 2edn. (Dover Publications, New York, 2004).

[18] S. Takahashi and S. Maekawa, Spin Hall Effect in Superconductors, *Jpn. J. Appl. Phys.* **51**(2012) 010110.

[19] D. Culcer and R. Winkler, Steady states of spin distributions in the presence of spin-orbit interactions *Phys. Rev. B* **76** (2007) 245322.

[20] R. Raimondi, C. Gorini, P. Schwab, and M. Dzierzawa, Quasiclassical approach to the spin Hall effect in the two-dimensional electron gas *Phys. Rev. B* **74** (2006) 035340.

[21] O. V. Dimitrova, Spin-Hall conductivity in a two-dimensional Rashba electron gas, *Phys. Rev. B* **71**(2005) 245327.

[22] J. I Inoue, G.E.W. Bauer, and L.W. Molenkamp, Diffuse transport and spin accumulation in a Rashba two-dimensional electron gas *Phys. Rev. B* **67** (2003) 033104.

[23] R. Raimondi and P. Schwab, Spin-Hall effect in a disordered two-dimensional electron system, *Phys. Rev. B* **71** (2005) 033311.

[24] C. Grimaldi, E. Cappelluti, and F. Marsiglio, Off-Fermi surface cancellation effects in spin-Hall conductivity of a two-dimensional Rashba electron gas, *Phys. Rev. B* **73** (2006) 081303.

[25] T. L. van den Berg, L. Raymond, and A. Verga, Dynamical spin Hall conductivity in a magnetic disordered system, *Phys. Rev. B* **84** (2011) 245210.

[26] H. Yavari, M. Mokhtari, and A. Bayervand, Temperature Dependence of the Spin-Hall Conductivity of a Two-Dimensional Impure Rashba Electron Gas in the Presence of Electron-Phonon and Electron-Electron Interactions, *J. Low. Temp. Phys.* **178** (2014) 331-44.

[27] H. Kontani, J. Goryo, and D.S. Hirashima, Intrinsic Spin Hall Effect in the s-Wave Superconducting State: Analysis of the Rashba Model, *Phys. Rev. Lett.* **102** (2009) 086602.

[28] S. Pandey, H. Kontani, D.S. Hirashima, R. Arita, and H. Aoki, Spin Hall effect in iron-based superconductors: A Dirac-point effect, *Phys. Rev. B* **86** (2012) 060507.

[29] K. V. Samokhin, Spin susceptibility of noncentrosymmetric superconductors, *Phys. Rev. B* **76** (2007) 094516.

[30] P. A. Frigeri, D.F. Agterberg, A. Koga, and M. Sigrist, Superconductivity without Inversion Symmetry: MnSi, *Phys. Rev. Lett.* **92** (2004) 097001.

[31] S. Fujimoto, Electron Correlation and Pairing States in Superconductors without Inversion Symmetry, *J. Phys. Soc. Jpn* **76** (2007) 051008.

[32] O. Dimitrova, and M. V. Feigel'man, Theory of a two-dimensional superconductor with broken inversion symmetry, *Phys. Rev. B* **76** (2007) 014522.

[33] G. Mahan, *Many-Particle Physics*, 3edn. (Plenum Press, New York, 2000).

[34] W. Pei, and L. You-Quan, The influence of inelastic relaxation time on intrinsic spin Hall effects in a disordered two-dimensional electron gas, *J. Phys.: Condens. Matter.* **20** (2008) 215206.